\documentclass[useAMS,usenatbib,usegraphicx]{mn2e}
\usepackage{amsmath,amssymb}
\usepackage{graphicx}
\usepackage{times}
\usepackage{natbib}
\usepackage{bm}
\usepackage{xcolor}

\title[]{The X-ray afterglow of GRB 081109A: clue to the wind bubble structure}
\author[]{Z. P. Jin$^{1,2}$, D. Xu$^{3}$\thanks{emails: jin@pmo.ac.cn (ZPJ) and dong@astro.ku.dk(DX)}, S. Covino$^{4}$, P. D'Avanzo$^{4}$, A. Antonelli$^{5}$, Y. Z. Fan$^{6,1}$,
D. M. Wei$^{1}$\\
$^1${\sl Purple Mountain Observatory, Chinese Academy of Sciences, Nanjing 210008, China}\\
$^2${\sl Graduate School, Chinese Academy of Sciences, Beijing, 100012, China}\\
$^3${\sl Dark Cosmology Centre, Niels Bohr Institute, University of Copenhagen, Juliane Maries Vej 30, DK-2100, Copenhagen, Denmark}\\
$^4${\sl INAF / Brera Astronomical Observatory, Via Bianchi 46, 23807, Merate (LC), Italy}\\
$^5${\sl INAF / Rome Astronomical Observatory, Via Frascati 33, 00044, Monte Porzio (Roma), Italy}\\
$^6${\sl Niels Bohr International Academy, Niels Bohr Institute, University of Copenhagen, Blegdamsvej 17, DK-2100 Copenhagen, Denmark}
}

\date{Accepted ......Received ......; in original form ......}

\begin{document}

\maketitle
\begin{abstract}
We present the prompt BAT and afterglow XRT data of {\em Swift}-discovered GRB 081109A up to $\sim
5\times 10^5$ sec after the trigger,
and the early ground-based optical follow-ups.
The temporal and spectral indices of the X-ray afterglow emission change remarkably.
We interpret this as the GRB jet first traversing the freely expanding supersonic stellar wind
of the progenitor with density varying as $\rho \propto r^{-2}$.
Then after approximately 300 seconds the jet traverses into a region of apparent constant density similar to that
expected in the stalled-wind region of a stellar wind bubble or the interstellar medium (ISM).
The optical afterglow data are generally consistent with such a scenario. Our best numerical model has
a wind density parameter {$A_{*} \sim 0.02$, a density of the stalled
wind $n\sim 0.12\,{\rm cm}^{-3}$, and a transition radius $ \sim 4.5 \times 10^{17}$ cm}. Such a
transition radius is smaller than that predicted by numerical simulations of the stellar wind
bubbles and may be due to a rapidly evolving wind of the progenitor close to the time of its
core-collapse.
 \end{abstract}



\section{Introduction}
Gamma-ray bursts (GRBs) are the most luminous explosions in the Universe.
They feature extremely relativistic outflows with bulk Lorentz factors of $10^{2-3}$ and isotropic energies of $10^{48-54}$ergs.
These energies are widely believed to be generated via the core-collapse of massive stars
for conventional long(-duration) GRBs (e.g., \citealt{Woosley93, Woosley06})
or the merger of compact star binaries for conventional short(-duration) GRBs (e.g., \citealt{Eichler89,Narayan92}).
In the standard fireball model the prompt soft $\gamma$-ray emission is powered by the collision of the material
shells within the relativistic outflow (i.e., the internal shocks);
afterwards these material shells spread and merge into a single uniform outflow, continuing to move outwards;
then the long-lived X-ray, optical and radio afterglow emission is powered by the interaction of the overall outflow and the
circum-burst medium (e.g., \citealt{Piran99,Meszaros02,Zhang04}).
Therefore, the temporal and spectral evolution of the multi-wavelength afterglow can be used to diagnose the underlying
radiation mechanism and the profile of the circumburst medium (e.g., \citealt{Sari98,Chevalier00,PK02}).

For short GRBs, the circum-burst medium is expected to be of the interstellar medium
(ISM) type or even of the intergalactic medium (IGM) type, for which the number density
is roughly constant and much lower than unity, being consistent with the current
afterglow modeling \citep[see][for a review]{Nakar07}.
For long GRBs, Panaitescu \& Kumar (2002) found that the afterglow modeling usually favors the constant density (CD) medium scenario,
half cases in 10 bursts of a sample were better fitted by the CD medium while only one better described by a free wind (FW) medium.
However, Starling et al. (2008) found that in 10 bursts of a sample, 4 were clearly in FW medium while only 1 consistent with CD medium.
It seems that the circum-burst environments are unlikely to be drawn from only one of the CD or FW profile for long bursts.
This is not obviously consistent with the collapsar model in which an ideal FW circum-burst medium would be created.
The FW has a density profile $\rho=5\times10^{11}A_{*}r^{-2}{\rm g~cm}^{-1}$,
here $A_{*}=(\frac{\dot{M}}{10^{-5}M_{\odot}~{\rm yr}^{-1}})(\frac{1000{\rm km~s}^{-1}}{v_{\rm w}})$ is the parameter reflecting the density of the wind.
One potential solution is that the ideal FW
medium profile has been modified before the GRB explosion. As is known, massive stars are
believed to enter the Wolf-Rayet stage during their late evolution and have lost a major
fraction of their masses in the form of the stellar wind. The interaction between this
stellar wind and the surrounding medium creates a bubble structure
(e.g.,\citealt{Weaver77,Wijers01,Ramirez01,Dai03,Chevalier04,vanm06,vanm07,vanm08,Eldridge06,Eldridge07,Peer06}). In this
scenario the free expanding supersonic wind is terminated at a radius $R_{\rm t}\sim 10^{18}-10^{20}$cm,
where the density jumps by a factor of 4 or more with the lower value expected for such an adiabatic shock.
Beyond the wind termination shock is the roughly constant density stalled wind,
holding up to a rather large radius $R_{\rm ISM}\sim 10^{19}-10^{21}$cm, outside this radius is the very dense swept-up ISM (the number density $n\sim 10^{2}-10^{3}~{\rm cm^{-3}}$) and then the ISM.

If the above picture is a good approximation of the circum-burst medium,
it may be possible to observed afterglow signatures caused by the medium transition at $\sim R_{\rm t}$ or even at $\sim R_{\rm ISM}$.
In Section 2, we describe the signatures in GRB afterglows for the first transition which is more easily
observed and more of interest with respect to the second one.
Such a transition is possibly evident in GRB\,050319 (Kamble et al. 2007) and GRB\,050904 (Gendre et al. 2007).
In Section 3, we present the X-ray afterglow observations of GRB\,081109A which leads to reach the conclusion
that the FW-CD transition is quite clear in this event.
The optical afterglow data are generally consistent with such a scenario.
Throughout this work we use the notation $F (t, \nu) \propto t^{\alpha} \nu^{\beta}$ for the
afterglow monochromatic flux as a function of time, where $\nu$ represents the observer's frequency,
$\alpha$ is the monochromatic flux decay index, and $\beta$ is the energy spectrum index. The convention
$Q_x=Q/10^x$ has been adopted in cgs units.

\section{The afterglow signatures of the medium transition}
The GRB afterglow emission in the CD or FW scenarios has been extensively discussed (see Zhang \& Meszaros 2004
for a review).
We use the standard fireball afterglow theory with the simple microphysical assumptions
of constant energy fractions imparted to the swept-up electrons, $\epsilon_{\rm e}$, and
to the generated magnetic field, $\epsilon_{\rm B}$, respectively. The typical synchrotron radiation
frequency $\nu_{\rm m}$ and the cooling frequency $\nu_{\rm c}$ are calculated in the standard way \citep{Sari98}.
Both scenarios lead to the afterglow
closure relations made of the temporal decay index $\alpha$ and the spectral index
$\beta$, depending upon the spectral segment and the energy distribution index of $p\sim
2-3$ of the swept-up electrons. Usually the synchrotron self-absorption frequency,
$\nu_{\rm a}$, is much lower than the optical band and thus neglected unless radio
observation invoked. Therefore, with good-quality temporal and spectral data, the
circum-burst medium profile can be reliably constrained. We summarize the closure
relation in Table~\ref{table:closure} apart from the $\nu<\nu_{\rm a}$ cases \citep[see also][]{Peer06}.

For the X-ray afterglow of our interest, there are two ways to diagnose the number density profile of the medium:\\(I) When
$\nu_{\rm m}<\nu_{\rm opt}<\nu_{\rm c}<\nu_{\rm X}$ occurs, in the FW scenario the X-ray emission
drops with time as $t^{(2-3p)/4}$ while the optical emission drops faster by a factor of $1/4$. On
the contrary, in the CD scenario the optical emission drops more slowly than the X-rays.\\(II) In the case of $\nu_{\rm
m}<\nu_{\rm X}<\nu_{\rm c}$, i.e., the X-ray emission is in the slow cooling
phase, the temporal and spectral indices roughly satisfy $\alpha=(3\beta-1)/2$ in the FW scenario.

Then how does the afterglow evolve after the FW-CD transition at the radius $R_{\rm t}$?
It is known that the characteristic frequency $\nu_{\rm m}$ decays with time as $t^{-3/2}$
in both CD and FW scenarios. However, the cooling frequency $\nu_{\rm c}$
evolves very differently. In the FW scenario it increases as $ t^{1/2}$ but in
the CD scenario it declines as $t^{-1/2}$. Therefore, the following observational
signatures would be evident before and after the transition. In Case I, the deceline of the
X-ray emission remains unchanged while the optical decay becomes shallower by a factor of
$t^{1/2}$ as long as the optical band is still below $\nu_{\rm c}$. In Case II, the X-ray
decay will get flattened by a factor of $t^{1/2}$ as long as the X-ray band is still below
$\nu_{\rm c}$. Afterwards, when $\nu_{\rm c}$ drops below the X-ray band, the X-ray decay
steepens by a factor of $t^{-1/4}$. As we'll show in next Section, the X-ray afterglow of
GRB\,081109A fits Case II.

\begin{table}
\caption{Temporal index $\alpha$ and spectral index $\beta$ in the cases that the medium
profile can be constrained, where the convention $F (t, \nu) \propto
t^{\alpha}\nu^{\beta}$ is adopted, $p$ is the power-law index of the swept-up electrons,
and $\nu_{\rm a}$, $\nu_{\rm c}$, and $\nu_{\rm m}$ are the self-absorption frequency,
the cooling frequency and the characteristic frequency in the synchrotron radiation,
respectively (\citealt{Sari98,Chevalier00}).}
\begin{center}
\begin{tabular}{c|c|c|c|c}

\hline

Case & $\beta$ & $\alpha$ (CD) & $\alpha$ (FW)  \\

\hline

$\nu_{\rm a}<\nu<\nu_{\rm c}<\nu_{\rm m}$ & $1/3$ & ${1/6}$ & ${-2/3}$  \\

$\nu_{\rm a}<\nu<\nu_{\rm m}<\nu_{\rm c}$ & $1/3$ & $1/2$ & $0$ \\

$\max\{\nu_{\rm a},\nu_{\rm m}\}<\nu<\nu_{\rm c}$ & $-\frac{p-1}{2}$ & $\frac{3-3p}{4}$ & $\frac{1-3p}{4}$  \\

$\nu_{\rm c}<\nu$ & $-\frac{p}{2}$  & $\frac{2-3p}{4}$  &  $\frac{2-3p}{4}$\\

\hline

\label{table:closure}
\end{tabular}
\end{center}
\end{table}

\section{The FW-CD transition in GRB\,081109A}

\subsection{Observation and data reduction of GRB\,081109A}
\begin{figure}
\begin{center}
\includegraphics[height=11cm, width=9cm]{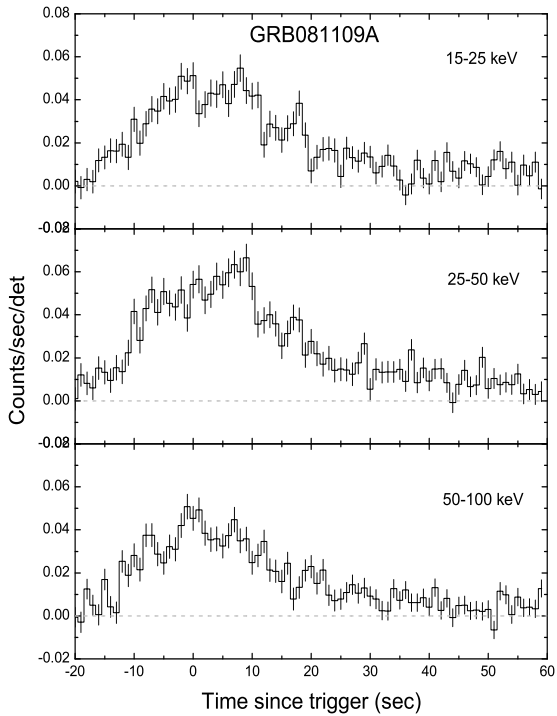}
\end{center}
\caption{The prompt lightcurves of GRB\,081109A in 15-25, 25-50, 50-100 keV with 1 s binning.
Spectral lag between different energy bands cannot be well measured for this event due to
relatively low signal to noise.}
\label{fig:T90}
\end{figure}

GRB\,081109A was triggered and located by Burst Alert Telescope (BAT) onboard {\em Swift} at
07:02:06 UT (trigger 334112), November 9th, 2008 (see \citealt{Immler08}). We downloaded the raw
data from the UK {\em Swift} data archive and processed them in a standard way with HEAsoft 6.6.
The BAT lightcurve in the 15-350 keV band was processed with the {\tt batgrbproduct} task. This
burst has a $T_{\rm 90}$ about 61 seconds in 15-350 keV, classified as a long GRB. The
duration value is consistent with the duration measurement of 45 s in 8-1000 keV by the Gamma-ray
Burst Monitor (GBM) onboard {\em Fermi} (\citealt{Kienlin08}). Also the time-averaged GBM spectrum
is best fit by a power-law function with an exponential high energy cutoff. The power law index
is -1.28$\pm$0.09 and the cutoff energy, parameterized as $E_{\rm p}$ is 240$\pm$60 keV (chi
squared 510 for 478 d.o.f.). The fluence in 8-1000keV is $6.35\pm0.43\times10^{-6}{\rm
erg}\cdot{\rm cm}^{-2}$. Fig.\,\ref{fig:T90} shows the prompt lightcurves of GRB\,081109A in 15-25,
25-50, 50-100 keV with 1 s time binning. Spectral lag between different energy bands cannot be well
measured for this event due to relatively low signal to noise.

\begin{figure}
\begin{center}
\includegraphics[height=7cm, width=9cm]{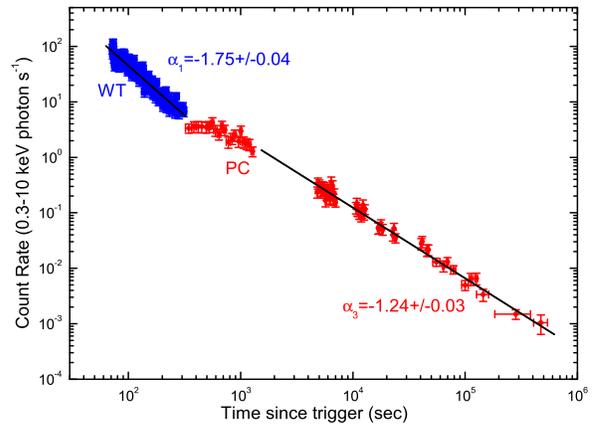}
\end{center}
\caption{The X-ray afterglow lightcurve of GRB\,081109A in 0.3-10 keV. Also marked are the fitting
temporal indices for the WT and late PC segments of the lightcurve.} \label{fig:XRT1}
\end{figure}

\begin{figure}
\begin{center}
\includegraphics[height=11cm, width=9cm]{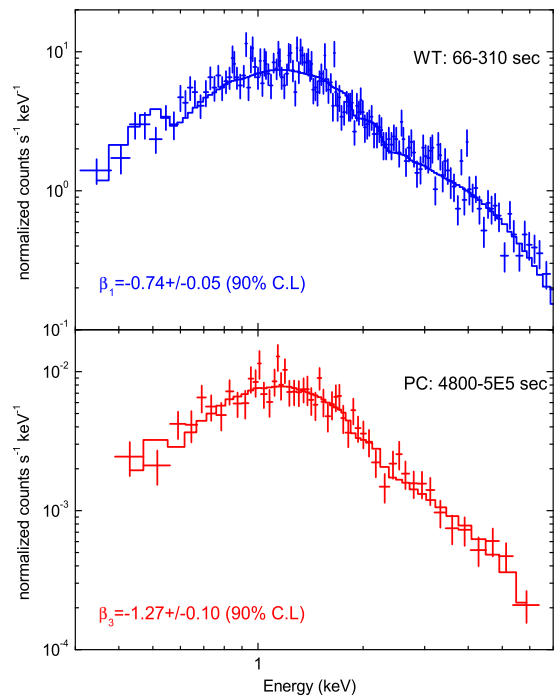}
\end{center}
\caption{The spectra of the WT and late PC segments of the X-ray afterglow of GRB\,081109A. An
absorbed power-law fitting is adopted. Also marked are the fitting spectral indices respectively.}
\label{fig:SED}
\end{figure}

The {\em Swift} X-ray Telescope (XRT) observations began at 65.6 seconds after the BAT trigger and
discovered a bright and fading X-ray afterglow. Observations continued during the following hours
and days and in several return visits, with Windowed Timing (WT) mode for $\sim 300$ s after the
trigger and Photon Counting (PC) mode afterwards. Throughout the X-ray observation, spectral
softening is evident. There is no need for pile-up correction to WT data as the highest count rate
is less than $\sim$150 count s$^{-1}$, but such correction should be applied to early PC data when
the count rate is higher than $\sim 0.6$ count s$^{-1}$ in order to get correct X-ray lightcurve
and spectra. We made this correction by fitting a King function profile to the point spread
function (PSF) to determine the radial point at which the measured PSF deviates from the model. The
counts were extracted using an annular aperture that excluded the affected $\sim$4 pixel core of
the PSF, and the count rate was corrected according to the model. We also considered the most
recent calibration and exposure maps. Fig.\,\ref{fig:XRT1} shows the 0.3-10 keV X-ray lightcurve,
which can be well modeled with a doubly broken power-law which is equivalent with a broken
power-law with a jump transition at $\sim$500 s. For a doubly broken power-law, the fitted
parameters are: $\alpha_1 = -1.75 \pm 0.04$ ($\chi^2=152.4$ for 119), $t_{\rm b1} \sim 310$ s,
$\alpha_2 = -0.70 \pm 0.13$ ($\chi^2=13.7$ for 18), $t_{\rm b2} \sim 2.9 \times 10^3$ s, $\alpha_3
= -1.24 \pm 0.03$ ($\chi^2=38.9$ for 54). The integrated spectra of the above first and third
segments are shown in Fig.\,\ref{fig:SED}. In detail, the spectral power-law indices are $\beta_1=
-0.74\pm0.05$, $\beta_3=-1.27\pm0.10$, and $\beta_2$ is in the middle.

The {\em Swift} Ultraviolet and Optical Telescope (UVOT) began observing at $\sim$150 s after
the trigger and found no optical counterpart down to $\sim$18 mag (\citealt{Immler08}).
Ground-based observations of the afterglow of GRB\,081109A were carried out with the REM telescope at La Silla
(\citealt{Zerbi01,Chincarini03,Covino04}) equipped with the ROSS optical spectrograph/imager and
the REMIR near-infrared camera on 2008 Nov 09, starting about 52 seconds after the burst
(\citealt{DAvanzo08}). The night was clear, with a seeing of about $2.0''$. We collected images
with typical exposures times from 10 to 120 seconds, covering a time interval of about 0.5 hours.
Image reduction was carried out by following the standard procedures.
Astrometry was performed using the
USNOB1.0\footnote{http://www.nofs.navy.mil/data/fchpix/} and the
2MASS\footnote{http://www.ipac.caltech.edu/2mass/} catalogues. We performed aperture photometry for
the afterglow and comparison stars. The afterglow was not detected in the optical. In the NIR it
was detected only in the $H$ and $Ks$ bands. Results for the photometry are reported in Table\,\ref{table:optical}.
The GROND instrument equipping the 2.2m ESO/MPI telescope at La Silla started observation 17.1 hr
after the trigger, gave a redshift limit of $z<3.5$ and a best fit of intrinsic extinction of Av
between 0.6 and 1.2 (\citealt{Clemens08}).

\begin{table}
\begin{center}
\caption{Optical photometry for GRB 081109A. Upper limits are at 3$\sigma$ confidence level.}
\begin{tabular}{c|c|c|c}

\hline

Filter & Time since trigger (s) & Time integral (s)  & Magnitude  \\

\hline

H & 89.4 & 71.3 & $>$15.0 \\
H & 168.9 & 72.2 & 15.017$\pm$0.115 \\
H & 248.4 & 71.3 & $>$15.4 \\
H & 327.0 & 71.3 & 14.893$\pm$0.109 \\
H & 394.4 & 47.3 & 14.747$\pm$0.128 \\
H & 636.3 & 71.3 & 14.963$\pm$0.114 \\
H & 990.6 & 121.1 & $>$15.5 \\
H & 1545.3 & 171.7 & 15.763$\pm$0.101 \\
K & 450.6 & 45.6 & 14.365$\pm$0.259 \\
K & 718.4 & 72.2 & 14.465$\pm$0.207 \\
K & 1122.8 & 121.1 & 14.645$\pm$0.188 \\

\hline

\label{table:optical}
\end{tabular}
\end{center}
\end{table}

\subsection{Analytical investigation of the afterglow of GRB 081109A}

For $t<t_{\rm b1}$, the temporal index $\alpha_1=-1.75\pm0.04$ and the spectral index
$\beta_1=-0.74\pm0.05$ are related by $\alpha-3\beta/2 \approx -0.5$, being consistent
with the forward-shock emission in the FW medium as long as $\nu_{\rm m}<\nu_{_{\rm
X}}<\nu_{\rm c}$. For $t>t_{\rm b2}$, the temporal index $\alpha_3=-1.24\pm0.03$ and the
spectral index $\beta_{3}=-1.27\pm0.10$ are related by $\alpha-3\beta/2\sim 0.5$,
suggesting that the medium can be either FW or CD given $\nu_{_{\rm X}}>\max\{\nu_{\rm
m},\nu_{\rm c}\}$ (\citealt{Zhang04}). If the FW scenario holds from the very
beginning, we have $\nu_{\rm c}(t_{\rm b2})>\nu_{\rm c}(t_{\rm b1})>\nu_{_{\rm X}}$
because of $\nu_{\rm c} \propto t^{1/2}$. One can, of course, assume that either  $A_{*}$ or
$\epsilon_{\rm B}$ has increased abruptly and then get a $\nu_{\rm c}(t_{\rm b2})\ll \nu_{\rm c}(t_{\rm
b1})$. However, such a treatment is lack of any solid physical background and is thus
artificial. On the other hand, if we assume that at $t\sim t_{\rm b1}$ there comes the FW to CD transition,
the forward shock emission light curve will get flattened by a factor of $t^{1/2}$, roughly
consistent with the data (please note that $\alpha_2$ is only poorly constrained). In this
scenario, the steepening at $t\geq t_{\rm b2}$ implies that $\nu_c(t_{\rm b2})>\nu_{_{\rm X}}$.
Below we present our quantitative estimates.

In a FW medium, we have $\alpha_1=\frac{1-3p}{4}$ and
$\beta_1=-\frac{p-1}{2}$ in the case of $\nu_{\rm m}<\nu_{\rm X}<\nu_{\rm c}$. In a CD medium, we have
$\alpha_3=\frac{2-3p}{4}$ and $\beta_3=-\frac{p}{2}$ for $\nu_{\rm X}>\max\{\nu_{\rm m},\nu_{\rm c}\}$.
We find that $p=2.5$ fits both the temporal and spectral slopes of GRB 081109A.

In a FW medium, we have \citep[e.g.,][]{Chevalier00}:
\begin{equation}
F_{\nu,{\rm max}} = 0.23~{\rm Jy}~D_{\rm L,28.34}^{-2}\epsilon_{\rm B,-2}^{1/2}E_{\rm
k,53}^{1/2}A_{*}t_{\rm d,-3}^{-1/2} , \label{eq:F_nu,max1}
\end{equation}
\begin{equation}
\nu_{\rm m} =1.8\times 10^{16}~{\rm Hz}~C_{p}^2(\frac{1+z}{2})^{1/2}\epsilon_{\rm e,-1}^2
\epsilon_{\rm B,-2}^{1/2} E_{\rm k,53}^{1/2}t_{\rm d,-3}^{-3/2},\label{eq:nu_m1}
\end{equation}
\begin{equation}
\nu_{\rm c} = 1.1\times 10^{13}~{\rm Hz}~(\frac{1+z}{2})^{-3/2}\epsilon_{\rm B,-2}^{-3/2}E_{\rm k,
53}^{1/2}A_{*}^{-2}
 t_{\rm d,-3}^{1/2},
 \label{eq:nu_c1}
 \end{equation}
where $E_{\rm k}$ is the isotropic equivalent energy, $z$ is the redshift of the GRB and $D_{\rm
L}$ is the corresponding luminosity distance, $t_{\rm d}$ is the time in days since trigger in the
observer's frame, and $C_{p}\equiv13(p-2)/[3(p-1)]$.

In our model, at  $t\sim 65.6$ s, $\nu_{\rm m}<0.3{\rm keV}$, $\nu_{\rm c}>10{\rm keV}$  and $F_{\rm
0.3keV}\sim2.6\times10^{-3}~{\rm Jy}$  are needed. So we have
\begin{equation}
\epsilon_{\rm e,-1}^{4}\epsilon_{\rm B,-2}E_{\rm k,53}<3~~~~~~(\nu_{\rm m}<0.3{\rm keV}),
\label{eq:wind,m,X}
 \end{equation}
\begin{equation}
\epsilon_{\rm B,-2}^{3/4}E_{\rm k,53}^{-1/4}A_{*}<0.002~~~(\nu_{\rm c}>10{\rm keV}),
\label{eq:wind,X,c}
 \end{equation}
\begin{equation}
\epsilon_{\rm e,-1}^{3/2}\epsilon_{\rm B,-2}^{7/8}E_{\rm k, 53}^{7/8}A_{*}\sim 0.01.
\label{eq:wind,FX}
\end{equation}

In a CD medium, we have \citep[e.g.,][]{Sari98}:
\begin{equation}
F_{\nu,{\rm max}}=6.6~{\rm mJy}~({1+z\over 2}) D_{\rm L,28.34}^{-2} \epsilon_{\rm
B,-2}^{1/2}E_{\rm k,53}n_0^{1/2}, \label{eq:F_nu,max2}
\end{equation}
\begin{equation}
\nu_{\rm m} =2.4\times 10^{16}~{\rm Hz}~C_p^2 ({1+z \over 2})^{1/2} \epsilon_{\rm
e,-1}^2\epsilon_{\rm B,-2}^{1/2} E_{\rm k,53}^{1/2} t_{\rm d,-3}^{-3/2},\label{eq:nu_m2}
\end{equation}
\begin{equation}
\nu_{\rm c} = 4.4\times 10^{16}~{\rm Hz}~({1+z \over 2})^{-1/2}\epsilon_{\rm B,-2}^{-3/2}E_{\rm k,
53}^{-1/2}n_0^{-1}
 t_{\rm d,-3}^{-1/2},
 \label{eq:nu_c2}
 \end{equation}
{where $n$ is the density of CD medium.}

At $t\sim t_{\rm b2}\sim 2900$s, our model suggests that $\nu_{\rm m}$ and $\nu_{\rm c}<0.3{\rm
keV}$, and $F_{\rm 0.3keV}\sim5.7\times10^{-5}~{\rm Jy}$. We then have

\begin{equation}
\epsilon_{\rm e,-1}^{4}\epsilon_{\rm B,-2}E_{\rm k, 53}<8\times10^4 ~~~(\nu_{\rm m}<0.3{\rm keV}),
\label{eq:ISM,m,X}
\end{equation}

\begin{equation}
\epsilon_{\rm B,-2}^{3}E_{\rm k, 53}n_0^{2}>0.01 ~~~(\nu_{\rm c}<0.3{\rm keV}), \label{eq:ISM,X,c}
\end{equation}

\begin{equation}
\epsilon_{\rm e,-1}^{3/2}\epsilon_{\rm B,-2}^{1/8}E_{\rm k, 53}^{9/8}\sim 1.8. \label{eq:ISM,FX}
\end{equation}

In a termination shock model, the crossing time is estimated by
\cite{Chevalier04}:
\begin{equation}
t(R_{\rm t})=1.5{\rm h}(\frac{1+z}{2})E_{\rm k,53}^{-1}A_{*,-1}^{2}n_{0}^{-1}\sim310{\rm s},
 \label{eq:R_t}
\end{equation}
i.e.,
\begin{equation}
E_{\rm k,53}^{-1}A_{*,-1}^{2}n_{0}^{-1}\sim0.06. \label{eq:R_t_2}
\end{equation}

For GRB 081109A, there is no self-consistent solution for
Eqs.(\ref{eq:wind,m,X}-\ref{eq:wind,FX}), Eqs.(\ref{eq:ISM,m,X}-\ref{eq:ISM,FX}) and
Eq.(\ref{eq:R_t_2}) provided that $\epsilon_{\rm B}$ is a constant in the free wind and in the CD medium.
With eqs.(\ref{eq:R_t_2}), (\ref{eq:wind,X,c}) and (\ref{eq:ISM,X,c}), we have
$\epsilon_{\rm B,CD}/\epsilon_{\rm B,w}\geq 7$, where the subscripts ``${\rm CD}$" and ``${\rm w}$"
represent the physical parameters measured in CD and FW medium, respectively.
A similar assumption was needed in the modeling of the afterglow data of GRB 050904 (\citealt{Gendre07}) and GRB050319 (\citealt{Kamble07}).
The physical reason is that the CD medium has been heated by the termination reverse shock and then may be weakly magnetized.

\begin{figure*}
\begin{center}
\includegraphics[angle=0, width=164mm]{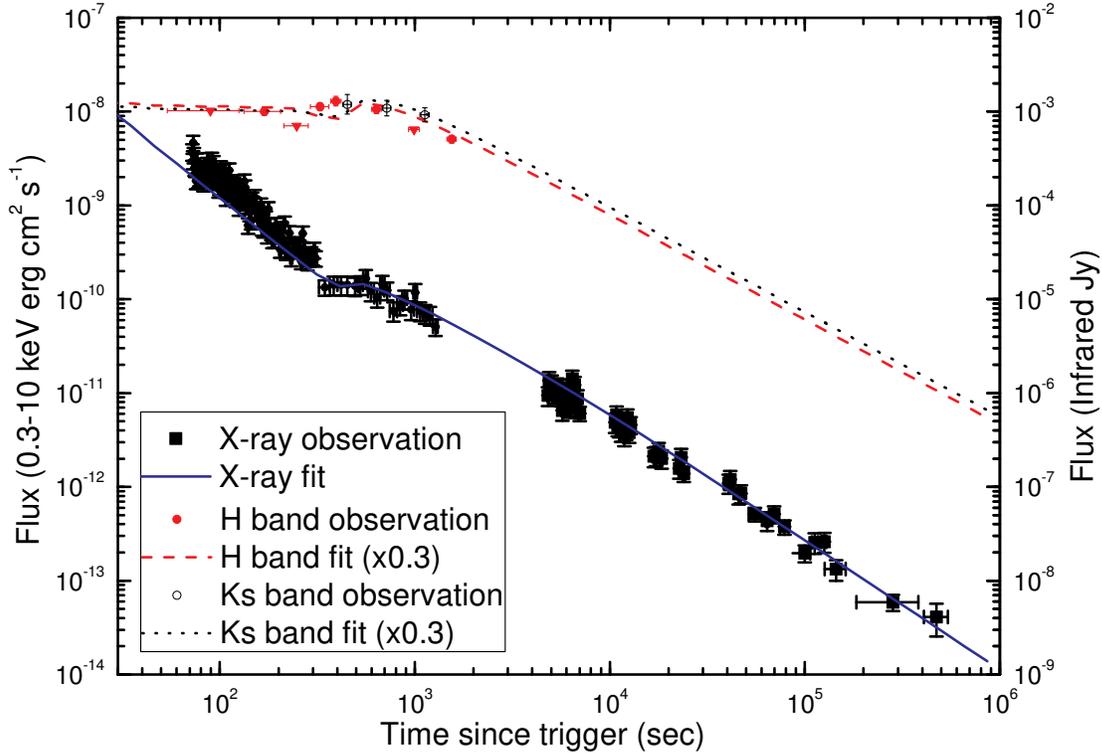}
\end{center}
\caption{Numerical fit to the afterglow of GRB081109A. The solid circles and squares
are X-ray observations, and the solid line is our numerical fit to X-ray. The
circles on the top are H band (empty) and Ks band (solid) observations, the
triangles are upper limits in H band. The dashed and dotted lines are numerical fit
to H band and Ks band data, respectively.} \label{fit}
\end{figure*}

The optical data, though rare comparing with the X-ray ones, provide us a reliable test of the
current afterglow model. As shown in Fig.\,\ref{fit}, the H-band lightcurve is distinguished by
 a flat at early time and then a re-brightening at $t\sim 400$ s. These features are generally consistent
 with our FW to CD medium model. In the FW medium, a flat segment is expected if the
 observer's frequency $\nu_{\rm obs}$ is above the synchrotron self-absorption frequency $\nu_{\rm a}$ but
 below $\nu_{\rm m}~(<\nu_{\rm c})$. In the CD medium, a re-brightening is present if
 the relation $\nu_{\rm a}<\nu_{\rm obs}<\nu_{\rm m}<\nu_{\rm c}$ still holds \citep[e.g.,][]{Zhang04}.
The re-brightening peaks when $\nu_{\rm m}$ crosses the observer's frequency. After then
the flux will drop with time as $t^{3(1-p)/4}$, a little bit shallower than the simultaneous X-ray decline,
as suggested by the H-band data.

\subsection{Numerical fit of the afterglow of GRB 081109A}
\begin{figure}
\begin{center}
\includegraphics[width=0.45\textwidth]{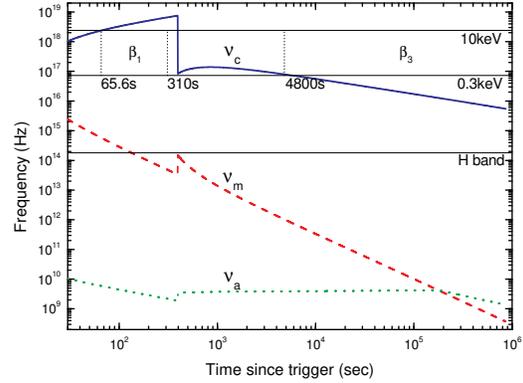}
\end{center}
\caption{Numerical fit to the spectral evolution in GRB081109A afterglow. In the
X-ray band(0.3-10keV), the spectrum between 65.6s and 310s is a single power law
$\beta_{1}=0.75$, and after $\sim4800$s is another single power law
$\beta_{3}=1.25$.} \label{info}
\end{figure}

The code used here to fit the X-ray light curves has been developed by \cite{Yan07}, with small
changes to adapt a density transition in surrounding medium. Assuming a FW to CD medium transition,
we find out that the observation data can be reasonably reproduced with the following parameters (see Fig.\,\ref{fit} and Fig.\,\ref{info}):
$E_{\rm k}=4\times10^{54}$erg, the initial
Lorentz factor $\gamma_{0}=500$, $A_{*}=0.02$, $n=0.12~{\rm cm^{-3}}$, $R_{\rm t}=4.5\times10^{17}$cm, $\epsilon_{\rm e}=0.02$,
$p=2.5$, $\epsilon_{\rm B,w}=0.0002$ and $\epsilon_{\rm B,CD}=0.001$. The source is assumed
to be at a redshift $z=1$. As shown in Fig.\,\ref{info},
a sudden increase of $\epsilon_{\rm B}$ for $t \geq t_{\rm b1}$ is required to account for the
jump of $\nu_{\rm c}$ inferred from the X-ray data.
$A_{*}\sim0.01$ and $R_{\rm t} \sim 10^{17}-10^{18}$ cm is lower than the typical value found in numerical simulations
and may be due to a rapidly evolving wind of the progenitor close to the time of its core-collapse (\citealt{Eldridge06,Eldridge07,vanm08}).
Assuming a redshift $z=1$, the isotropic energy $E_\gamma$ in the energy range of $1-10000$ keV is about $5\times10^{52}$ergs. The
 corresponding GRB efficiency $E_\gamma/(E_\gamma+E_{\rm k})$ is $\sim 1\%$. Such a low efficiency, though
 not typical, is still reasonable, as found in previous estimates \citep[e.g.,][]{FP06,Jin07}.

The observed optical flux is lower than the value extrapolated from X-ray observation, which implies that there is extinction in optical band.
To agree with the optical observation, our numerical fit for the $H$ and $Ks$ band has been adjusted by a factor of 0.3,
which requires about 1 mag extinction caused by the GRB host galaxy at the observed $H$ and $Ks$ band. They are about 1 micron at the GRB host galaxy.
It essentially means E$_{\rm B-V} \sim 0.8-0.9$ assuming Milky Way or SMC extinction curves.
With a Galactic dust to gas ratio it would correspond to N$_{\rm H} \sim 5\times10^{21}$ cm$^{-2}$.
The fit is shown in Fig.\,\ref{fit}.
The small differences around the termination radius may due to the sharp density jump we have assumed.

Before the blast wave reaches $R_{\rm t}$, its radius can be estimated as (\citealt{Chevalier00}):
\begin{equation}
R=3.5\times10^{17}(\frac{1+z}{2})^{-1/2}E_{\rm k,53}^{1/2}A_{*}^{-1/2}t_{\rm d}^{1/2}~{\rm cm}. \label{eq:wind,R}
\end{equation}
For $t(R_{\rm t}) \sim 310$ s, with eq.(\ref{eq:wind,X,c}) we have $R=R_{\rm t} \geq 10^{22}[(1+z)/2]^{-1/2}A_{*}^{3/2}\epsilon_{\rm B,-2}^{3/2}~{\rm cm}$.
{\it This implied that $R_{\rm t}$ depends on the undetermined redshift $z$ weakly}. A larger $A_*$ requires a
smaller $\epsilon_{\rm B,w}$ otherwise $R_{\rm t}$ will be much smaller than a few $\times 10^{17}$ cm,
the lowest value expected in the numerical simulation. Since the derived $\epsilon_{\rm B,w}$ is already
as low as $\sim  10^{-4}$, a smaller value is less likely. That is why we will not consider
the case of a FW parameter $A_{*}\gg 0.01$.

When a GRB jet finally enters into the very dense swept-ISM region, strong reverse shock may be
formed and an afterglow re-brightening is expected \citep{Dai03,Peer06}. The X-ray observation for
GRB081109A lasted to about $4.7\times 10^{5}$ s after the trigger and did not find obvious flux
enhancement. In our numerical fit, the jet front reached a radius $\sim 1.9\times10^{18}$cm at such
a late time, much smaller than $R_{\rm ISM}$ estimated by numerical simulations
(\citealt{Eldridge06,Eldridge07}). So the GRB outflow was still in the shocked wind material
ejected in the Wolf-Rayet stage with an approximately constant density (\citealt{Weaver77}).

\section{Conclusion and discussion}
Although some GRBs are inferred to occur in free stellar wind medium \citep[e.g.,][]{PK02,Starling08},
most are inferred to occur in a medium with a constant number density \cite[even for some bursts associated with
bright supernovae, see ][and the references therein]{Fan08},
which may indicate that GRB outflows expand into the wind bubble
rather than the ideal free stellar wind. As shown in Section 2 of
this work, in some cases the X-ray afterglow data could shed light
on the wind bubble structure. Therefore the X-ray afterglow observation since the
early time is required to trace the profile of the circum-burst
medium. However, in many {\em Swift} GRB events the early X-ray
afterglow deviates from the standard afterglow model significantly
(\citealt{Nousek06,Zhang06}). For instance, the prolonged activity
of GRB central engines would generate energetic X-ray flares that
have outshone the regular forward shock emission (e.g.,
\citealt{Fan05,Zhang06,Nousek06}). Fortunately, in GRB\,081109A
there is no flare accompanying the early X-ray afterglow. The
temporal and spectral evolutions of the X-ray afterglow imply a
medium transition at the radius $R_{\rm t} \sim 4.5\times 10^{17}$ cm
(see Section 3). Such a small $R_{\rm t}$ implies a small wind
parameter $A_{*}$ and a large $p/k$ since
\[
 R_{\rm t} = 5.7 \times 10^{17} ~
({v_{\rm w} \over 10^3~{\rm km ~s^{-1}}}) ({p/k\over 10^6 {\rm cm^3 ~
K}})^{-1/2}A_{*,-2}^{1/2}~{\rm cm},
\]
where $p$ is the pressure in the shocked wind and $k$ is the Boltzmann constant
(\citealt{Chevalier04}). Indeed $A_{*}$ is found to be as small as $10^{-2}$
in our numerical fit. GRB\,081109A is thus a good candidate of long GRBs born in
wind bubble.

The rising behavior of the very early afterglow lightcurves can play an important role in probing
the density profile of the medium. This is particularly the case if the reverse shock optical
emission is very weak. As shown in \cite{Jin07} and \cite{xue09}, in the free wind scenario the
very early optical rise is usually not expected to be faster than $t^{1/2}$ while in the constant
density medium the rise can be faster than $t^2$. Therefore the $\sim t^3$-like rise in the early
optical/infrared/X-ray afterglows of GRB\,060418, GRB\,060607A (\citealt{Molinari07}), GRB\,060801,
GRB\,060926, GRB\,080319C, and GRB\,080413B (\citealt{xue09}) rules out the free wind medium for
$R>10^{16}$ cm. Thus the absence of the free wind signature in some GRB afterglows is still a
puzzle. One possible solution is that the mass loss rate due to the stellar wind before massive
stars collapse is not a constant and might be much lower than previously assumed
{(\citealt{Eldridge06,Eldridge07}).}

\section*{Acknowledgments}
We would like to thank the referee for the constructive comments and for the help in
improving the presentation of this work. This work is supported by the National
Science Foundation (grants 10673034 and 10621303) and National Basic Research
Program (973 programs 2007CB815404 and 2009CB824800) of China. DX is at the Dark
Cosmology Centre funded by Danish National Research Foundation. YZF is also
supported by Danish National Research Foundation and by Chinese Academy of Sciences.


\begin{thebibliography}{99}
\bibitem[Chevalier \& Li (2000)]{Chevalier00} Chevalier, R. A., Li, Z. Y., 2000, ApJ, 536, 195
\bibitem[Chevalier, Li \& Fransson (2004)]{Chevalier04} Chevalier, R. A., Li, Z. Y., Fransson, C., 2004, ApJ, 606, 369
\bibitem[Chincarini et al(2003)]{Chincarini03} Chincarini, G., Zerbi, F. M., Antonelli, A. et al. 2003, The Messenger, 113, 40
\bibitem[Clemens, Kruehler, \& Greiner (2008)]{Clemens08} Clemens, C., Kruehler, T., Greiner, J., 2008, GCN Circ., 8515
\bibitem[Covino et al. (2004)]{Covino04} Covino, S., Stefanon, M. Fernandez-Soto, A. et al. 2004, SPIE 5492, 1613
\bibitem[D'Avanzo et al. (2008)]{DAvanzo08} D'Avanzo, P., Covino, S., Antonelli, L.A. et al. 2008, GCN 8501
\bibitem[Dai \& Wu (2003)]{Dai03} Dai, Z. G., Wu, X. F., 2003, ApJ, 591, L21
\bibitem[Eichler et al. (1989)]{Eichler89} Eichler, D., Livio, M., Piran, T., Schramm, D. N., 1989, Nat, 340, 126
\bibitem[Eldridge et al. (2006)]{Eldridge06} Eldridge, J. J. et al., 2006, MNRAS, 367, 186
\bibitem[Eldridge (2007)]{Eldridge07} Eldridge, J. J., 2007, MNRAS, 377, L29
\bibitem[Fan (2008)]{Fan08} Fan Y. Z., 2008, MNRAS, 389, 1306
\bibitem[Fan \& Piran (2006)]{FP06} Fan Y. Z., Piran T., 2006, MNRAS, 369, 197
\bibitem[Fan \& Wei (2005)]{Fan05} Fan Y. Z., Wei D. M., 2005, MNRAS, 364, L42
\bibitem[Gendre et al. (2007)]{Gendre07} Gendre, B. et. al., 2007, A\&A, 462, 565
\bibitem[Immler et al. (2008)]{Immler08} Immler, S. et al., 2008, GCN Circ., 8500
\bibitem[Jin \& Fan (2007)]{Jin07} Jin Z. P., Fan Y. Z., 2007, MNRAS, 378, 1043
\bibitem[Kamble et al. (2007)]{Kamble07} Kamble, A., Resmi, L., Misra, K., 2007, ApJ, 664, L5
\bibitem[M\'{e}sz\'{a}ros (2002)]{Meszaros02} M\'{e}sz\'{a}ros, P., 2002, ARA\&A, 40, 137
\bibitem[Molinari et al. (2007)]{Molinari07} Molinari, E. et al., 2007, A\&A, 469, L13
\bibitem[Nakar (2007)]{Nakar07} Nakar, E., 2007, Phys. Rep., 442, 166
\bibitem[Narayan, Paczy\'{n}ski, \& Prian (1992)]{Narayan92} Narayan, R., Paczy\'{n}ski, B., Piran, T., 1992, ApJ, 395, L83
\bibitem[Nousek et al. (2006)]{Nousek06} Nousek, J. A. et al., 2006, ApJ, 642, 389
\bibitem[Panaitescu \& Kumar (2002)]{PK02} Panaitescu, A., Kumar, P., 2002, ApJ, 571, 779
\bibitem[Pe'er \& Wijers (2006)]{Peer06} Pe'er, A. \& Wijers, R. A. M. J., 2006, ApJ, 643, 1036
\bibitem[Piran (1999)]{Piran99} Piran, T., 1999, Phys. Rep., 314, 575
\bibitem[Ramirez-Ruiz et al. (2001)]{Ramirez01} Ramirez-Ruiz, E., Dray, L. M., Madau, P., Tout, C. A., 2001, MNRAS, 327, 829
\bibitem[Sari, Piran, \& Narayan (1998)]{Sari98} Sari, R., Piran, T.,  Narayan, R., 1998, ApJ, 497, L17
\bibitem[Starling et al. (2008)]{Starling08} Starling et al., 2008, ApJ, 672, 433
\bibitem[van Marle et al. (2006)]{vanm06} van Marle, A. J., Langer, N., Achterberg, A., Garcia-Segura, G., 2006, A\&A, 460, 105
\bibitem[van Marle et al. (2007)]{vanm07} van Marle, A. J., Langer, N., Garc\'{\i}a-Segura, G., 2007, A\&A, 469, 941
\bibitem[van Marle et al. (2008)]{vanm08} van Marle, A. J., Langer, N., Yoon, S.-C., Garc\'{\i}a-Segura, G., 2008, A\&A, 478, 769
\bibitem[von Kienlin et al. (2008)]{Kienlin08} von Kienlin, A., 2008, GCN Circ. 8505
\bibitem[Weaver et al. (1977)]{Weaver77} Weaver, R., McCray, R., Castor, J., Shapiro, P., Moore, R., 1977, ApJ, 218, 377
\bibitem[Wijers (2001)]{Wijers01} Wijers, R. A. M. J., 2001, Gamma-Ray Bursts in the Afterglow Era:Proceedings of the International Workshop Held in Rome, Italy, 17-20 October 2000, ESO ASTROPHYSICS SYMPOSIA. Edited by E. Costa, F. Frontera, and J. Hjorth. Springer-Verlag,306
\bibitem[Woosley (1993)]{Woosley93} Woosley, S., 1993, ApJ, 405, 273
\bibitem[Woosley \& Bloom (2006)]{Woosley06} Woosley, S. E., Bloom, J. S., 2006, ARA\&A, 44, 507
\bibitem[Xue et al. (2009)]{xue09} Xue, R. R., Fan, Y. Z., Wei, D. M., 2009, A\&A, 498, 671
\bibitem[Yan, Wei \& Fan (2007)]{Yan07} Yan, T., Wei, D. M., Fan, Y. Z., 2007, Chin. J. Astron. Astrophys., 7, 777
\bibitem[Zerbi et al(2001)]{Zerbi01} Zerbi, F. M., Chincarini, G., Ghisellini, G. et al. 2001, AN 322, 275
\bibitem[Zhang et al. (2006)]{Zhang06} Zhang, B. et al. 2006, ApJ, 642, 354
\bibitem[Zhang \& M\'{e}sz\'{a}ros (2004)]{Zhang04} Zhang, B., M\'{e}sz\'{a}ros, P., 2004, IJMPA, 19, 2385
\end{thebibliography}
\end{document}